\documentclass{emulateapj}

\usepackage{natbib}
\usepackage{graphicx}
\usepackage{amssymb,ulem}

\usepackage[dvipsnames]{color}

\shorttitle{Conditions for SN driven galactic wind}
\shortauthors{Nath, Shchekinov}

\begin{document}


\newcommand{\3}{\ss}
\newcommand{\n}{\noindent}
\newcommand{\eps}{\varepsilon}
\def\be{\begin{equation}}
\def\ee{\end{equation}}
\def\bea{\begin{eqnarray}}
\def\eea{\end{eqnarray}}
\def\de{\partial}
\def\msun{M_\odot}
\def\div{\nabla\cdot}
\def\grad{\nabla}
\def\rot{\nabla\times}
\def\ltsima{$\; \buildrel < \over \sim \;$}
\def\simlt{\lower.5ex\hbox{\ltsima}}
\def\gtsima{$\; \buildrel > \over \sim \;$}
\def\simgt{\lower.5ex\hbox{\gtsima}}
\def\etal{{et al.\ }}
\def\red{\textcolor{red}} 
\def\blue{\textcolor{blue}}


\title{CONDITIONS FOR SUPERNOVAE DRIVEN GALACTIC WINDS}

\author{Biman B. Nath$^1$, Yuri Shchekinov$^2$}
\affil{$^1$ Raman Research Institute, Sadashiva Nagar, Bangalore 560080, India}
\affil{$^2$ Department of Physics, Southern Federal University,
        Rostov on Don, 344090 Russia}
\email{biman@rri.res.in; yus@sfedu.ru}

\begin{abstract}
We point out that the commonly assumed condition for galactic outflows, that supernovae (SNe) heating is efficient in the central regions of starburst galaxies, suffers from invalid assumptions. We show that a large filling factor 
of hot ($\ge 10^6$ K) gas is difficult to achieve through SNe heating, irrespective of the  initial gas
temperature and density, and of its being uniform or clumpy.
We instead suggest that correlated supernovae from OB associations in molecular clouds in the central region can drive powerful outflows if the molecular surface density is  $> 10^3$ M$_{\odot}$ pc$^{-2}$.
\end{abstract}


\section{Introduction}
Standard models of supernovae driven galactic outflows assume a central region where the SNe
energy input is thermalized (e.g.,  \cite{larson74, clegg85, heckman90, suchkov94, strickland09, sharma13}). These models
posit that SNe explosions shock heat the interstellar medium (ISM) within this region (of
size $\sim 200$ pc); numerical simulations for winds also implement this assumption (e.g., \cite{suchkov94, suchkov96, cooper09}). The basic assumption is that even with a small heating efficiency ($\sim 0.1$), SNe can heat
a low density gas to
$\simgt 10^6$ K and launch a galactic wind.

There are two assumptions here, one involving the energetics of SNe explosions and another to do with the thermalization of this energy. While the energy budget  can be met in the case of high
supernova rate in starbursts, the process of thermalization assumes that SNe remnants overlap and reach a porosity larger than unity. \cite{suchkov96, strickland09}  have discussed this issue in the context of the observed X-ray emission from the outflowing hot gas in M82. The required SNe heating efficiency of $\epsilon$ is connected with the mass loading factor,  the ratio $\beta$ between the total mass deposition rate and the mass lost through SNe and stellar winds. The temperature and brightness of the gas depend on different combinations of $\epsilon$ and $\beta$, and \cite{strickland09} suggested an optimum condition of $\beta\sim 1\hbox{--}3$ and $\epsilon \sim 0.1\hbox{--}0.3$. They suggested that $\epsilon$ could be large in the case of a low density gas ($\sim 0.1$ cm$^{-3}$). The average density of the diffuse ionized medium in starbursts is however $\approx 24$ cm$^{-3}$ \citep{armus89}.

The question is whether or not SNe remnants can overlap in these regions and sufficiently heat the gas \citep{melioli04}. We argue in this paper that it is difficult to achieve high porosity for hot ($\simgt 10^6\hbox{--}10^7$ K) gas irrespective of the ambient density being small or large, if  SNe remnants
occur randomly in this region. In order to overcome this difficulty we suggest that galaxies 
approaching a galactic wind stage are likely to produce super star clusters with an enhanced star formation rate. We note here in passing that other processes have also been invoked to aid galactic winds, such as radiation pressure \citep{nath09, murray11} and turbulence \citep{scannapieco13}, and also cosmic rays \citep{uhlig12} which can operate outside the central region.

\section{Porosity in a uniform ISM}
Consider the estimation of the porosity of hot ($\simgt 10^6$ K) gas in a uniform ISM of ambient density $n_0$. In the context of the three-phase ISM model, the porosity of the coronal gas is estimated by the final volume of SN remnants when the shells decelerate to the sound speed of the ambient gas \citep{mckee77, cox05}.
The `hot' interior gas at this stage has a temperature $\sim 5 \times 10^5$ K, which was used to infer the three-phase model of the ISM. The same expression was, however, used by \cite{heckman90}  in order to derive a high value of porosity of a {\it hotter} gas at $\simgt 10^6$ K gas (for $n_0\sim 100$ cm$^{-3}$ and $T_0\sim 10^5$ K; their eqn 2)\footnote{We note that the \cite{slavin93} prescription would have yielded a value of porosity smaller by a factor $\sim 5$.}, and
this argument has been repeated by other authors (e.g., \cite{suchkov94, strickland09}). The average temperature of the interior gas at this shell speed is, however, less than $10^6$ K, and {\it cannot} be used to determine the porosity of gas with $T\simgt 10^6$ K. 

We recall that the average gas temperature inside a SNe remnant decreases rapidly after the gas cools down to a temperature $\sim 10^6$ K.  Cox (1972) showed that the energy of the remnant scales as $E(t) \propto R^{-2}$, after the shell enters the radiative phase (at shell radius $R_c$). One has, 
\be
E(t)= 0.22 \, E_0\, \Bigl ({R\over R_c} \Bigr )^{-2} \,, R_c= 22.3 \, {\rm pc} \,E_{51}^{5/17} \, n_0^{-7/17} \,,
\ee
where the initial explosion energy is $E_0=10^{51} E_{51}$ erg, and $n_0$ is the ambient particle density (in cm$^{-3}$). The corresponding timescale is $t_c \approx 5 \times 10^4 \, {\rm yr}\, E_{51}^{4/17}\, n_0^{-9/17}$. Inverting this relation, we have the shell radius at a time when the internal energy has decreased to a fraction $f$ of the initial value as $R(f)=0.47 \, R_c \, f^{-1/2}$.
The corresponding time scale is $t(f)=0.07 \, t_c \, f^{-7/4}$.
Denoting the supernova rate density as $\nu_{SN}$ (yr $^{-1}$ pc$^{-3}$), we can define the porosity
when the energy has decreased by a fraction $f$, as,
\be
P(f) \approx 1.6 \times 10^8 \, (f/0.5)^{-13/4} \, \nu_{\rm SN} \, E_{51}^{19/17} \, n_0^{-30/17} \,.
\label{eq:por}
\ee
We have scaled the expression to $f=0.5$, since  the average interior temperature falls to $\sim 10^6$ K at this stage,  according to \cite{cox72} (see his Figure 2a). This is borne out by the simulations of Shelton (1998) for $n_0=0.01$, where the interior gas temperature
decreases below $10^6$ K after $t_c=5 \times 10^5$ yr appropriate for this density, as expected.

The condition for $\sim 10^7$ K gas to overlap (required to explain X-ray observations ) 
is more stringent and therefore we can use the porosity in eqn \ref{eq:por} as an upper limit. Note that the shell speed in \cite{cox72} at this stage ($f=0.5$) drops down to $\sim 100$ km s$^{-1}$. 
We can also use this velocity criterion to estimate the porosity for hot gas. 

The typical supernova rate density in starburst regions $\nu_{\rm SN}
\sim 10^{-9}$ yr$^{-1}$ pc$^{-3}$, implies a porosity $\approx 0.2$ for $n_0=1$, and
can be even lower for the average density in starbursts \citep{armus89}. 
For $f\equiv \epsilon=0.1$, as suggested by \cite{strickland09}, the porosity is $\approx 0.3 \, n_0^{-30/17}$, still less than unity. Moreover, for $f\sim 0.1$ the temperature inside SN remnants decrease to $\le 3 \times 10^5$ K, not enough to explain the observed X-ray emission. Furthermore, if the gas reservoir of  $\sim 5 \times 10^7$ M$_{\odot}$ required to explain the mass loading rate in  M82 \citep{suchkov96},
 were uniformly distributed in the $200$ pc central region, the density would be $\approx 70$ cm$^{-3}$, precluding the possibility of a large porosity factor for hot gas. And if this gas were uniformly heated to $10^7$ K, then it should emit hard X-rays with $10^{44}$ erg s$^{-1}$, much larger than 
the observed $3 \times 10^{40}$ erg s$^{-1}$.

One could argue that most of the mass is in clouds with a small filling factor, which are destroyed by  shock waves. 
\cite{thornton98} showed that 
gas with $T \sim 10^7$ K 
occupies a fraction $(1/5)^3\sim 10^{-2}$ of the SN remnant volume (their Figs 5,6) until the radiative stage $t_c \sim 10^5$ yr, when the average temperature is $\sim 10^6$ K. Therefore even if the shells overlap with the interior gas at mostly $10^6$ K, the filling factor of $10^7$ K gas is small, of order $0.01$, and the probability of evaporating a cloud and heating it up to $10^7$ K (as required by X-ray observation) is of the same order. Therefore the clouds should have a {\it large, and not small} filling factor in order to explain the X-ray observations.

One could argue that SN remnants propagate mostly through a hot, low density medium.  For a $10^6$ K gas, with (isothermal) sound speed $\sim 100$ km s$^{-1}$, 
the remnants will stop expanding after reaching this speed. 
The porosity of hot SN bubbles will therefore still be given by $P\approx 0.02$ for $n_0=1$ cm$^{-3}$ and a SN rate density of few $\times 10^{-9}$ yr$^{-1}$ pc$^{-3}$, because it corresponds to the same shell speed. This is the porosity provided by new SNe remnants. However, a $10^6$ K gas will cool in a time scale of $10^5 n_0^{-1}$ yr, and will need new SN remnants with large filling factor to maintain its temperature, though the above estimate 
shows the filling factor to be small. {\it Therefore the porosity of hot gas is less than unity 
 irrespective of the density and temperature of the gas in which SNe explode.}

In general it is difficult to simultaneously satisfy the conditions
for vigorous star formation rate (SFR) and also fill the central region with hot gas.
We can use the Schmidt-Kennicutt law (Kennicutt 1998) $\Sigma_{\rm SFR} \approx 2.5 \times 10^{-4} \Sigma^{1.4}$, 
where $\Sigma_{\rm SFR}$ is in the units of M$_{\odot}$ yr$^{-1}$ kpc$^{-2}$ and $\Sigma$ in units of M$_{\odot}$ pc$^{-2}$. For a initial mass function (IMF) of \cite{kroupa01, chabrier03}, the corresponding supernova rate is $\sim 0.01$ yr$^{-1}$ (SFR/M$_{\odot}$ yr$^{-1}$). For a scale height $H$ pc, the supernova rate density is $
\nu_{\rm SN} \approx 2.5 \times 10^{-12} \, \Sigma^{1.4} \, H^{-1}$. 
The gas density for a uniform medium is $n_0=\Sigma/(2 H \mu m_H) \sim 15 \, \Sigma H^{-1}$ cm$^{-3}$.
Therefore the porosity for the $10^6$ K gas is given by (from eqn \ref{eq:por}),
\be
P  \approx
3 \times 10^{-6} \, \Sigma^{-0.365} \, H^{0.765} \, E_{51}^{19/17} \,.
\ee
The requirement that $P \ge 1$ gives 
\be
\Sigma \le 10^{-15} H^{2.1} \,.
\ee
Using the Silk-Elmegreen law of star formation (Silk 1997, Elmegreen 1987
), $\Sigma_{\rm SFR}\approx 0.017 \, \Sigma \, \Omega$, where $\Omega$ is the angular velocity in the units of Myr$^{-1}$, one gets a similar condition: 
$\Sigma \le 10^{-4} \, H \, \Omega^{1/0.765}$.
{\it Since a large density is required for high SFR, while a low density is required to achieve high porosity of hot gas, the combination of these two requirement leads to a condition that is difficult to meet.}

\section{Coherency of supernovae}
The problem boils down to maintaining a large porosity for high temperature gas with a given 
star formation, or equivalently, SNe rate. This can be achieved by allowing SNe remnants to explode in a coherent manner so as to avoid excessive radiative cooling and  act collectively. 
One has to invoke spatial and temporal clustering of SNe events in order to increase the efficiency of SN heating, even if the heating is confined to small region 
We can define the coherency of SNe by requiring the four-volume at $t_c$ 
to be of order unity ($4 \pi R_c^3 \, t_c \, \nu_{SN}/3=1$) in a given localized region. This will ensure that new SNe will help compensate for the radiative loss in this region and keep the interior gas at $\simgt 10^6$ K.

Massive stars (the progenitors of SNe) are likely to form in OB associations in molecular clouds (MCs). These sites can provide the spatial and temporal coherency needed for this scenario. In order for these SNe to emerge from the parent MC and then heat up the intercloud medium, it is necessary for the SNe explosions to destroy the MC. 

Consider a MC of mass $M_{\rm MC}=10^5 \, M_{\rm MC,5}\, M_{\odot}$ and radius $R_{\rm MC}=5 \, R_{\rm MC, 5}$ pc. One fiducial example of such a MC is provided by G0.253+0.016 (called the `Brick') in the central region of Milky Way, with a mass $\sim 1.3 \times 10^5$ M$_{\odot}$ and radius $\sim 2.8$ pc \citep{longmore12}. The density of molecular hydrogen ($\mu=2.3$) is $n= 5 \times 10^3 \, {\rm cm}^{-3} \, M_{\rm MC, 5} \, R_{\rm MC, 5}^{-3}$. The corresponding free-fall time scale
is $t_{\rm ff}\approx 2 \times 10^{15}  \, n^{-1/2} \, {\rm s} \approx 1 \, {\rm Myr} \, M_{\rm MC, 5} ^{-1/2} \, R_{\rm MC, 5}^{3/2}$. In this environment, a SN remnant shell speed will become $v_c=100$ km s$^{-1}$ at a length scale 
$R_c\approx (E/\rho)^{1/3} \, (2/5v_c)^{2/3}= 23 \, {\rm pc} \, (E_{51}/n)^{1/3}$, at $t_c= (2/5) R_c/v_c= 0.6 \, {\rm Myr} \, (E_{51}/n)^{1/3}$,  
 consistent with the results of 
\cite{chevalier99} for SNe inside MCs.

One can show that for a star formation rate $\epsilon_\ast M_{\rm MC}/t_{\rm ff}$, where
$\epsilon_\ast\approx 0.02$ is the star formation efficiency per free-fall time \citep{lada10}, the porosity is given by $P= (4 \pi/3) R_c^3 t_c \nu_{\rm SN} \approx 0.06 \,  n^{1/6}$, (similar to estimates in \S 2, since this is essentially the Silk-Elmegreen law applied at a small scale).
Although the porosity is formally less than unity in this case, the effect of even a single supernova can be catastrophic for a small MC, since $R_c\gg R_{\rm MC}$. 
This effect is expedited by the ionization of the MC by massive stars prior to SNe events \citep{monaco04}. \cite{walch12} have found from simulations of a single SN in a MC that

the ensuing ionization front disperses the cloud on a time scale comparable to the sound crossing time of the ionized gas, of order a Myr; moreover radiation pressure may help in this regard \citep{murray11}.

\section{Initiating a galactic wind}
Although star formation in MCs help to concentrate the effects of SN remnants, it remains to be seen whether the energetics are sufficient for galactic wind.
In the case of correlated multiple SNe, superbubbles can be energetic enough to break out of the disc and reach a sufficient height above the disc with enough momentum so as to seed a galactic wind \citep{maclow88, roy13} (and references therein).
The threshold mechanical luminosity for a superbubble to breakout with a Mach number of $5\hbox{--}10$ depends on the ambient density ($\rho_0$) and sound speed ($c_s$), and the scale height ($H$). \cite{roy13} has determined 
a threshold luminosity of $L_{cr}\approx 100 \rho_0 H^2 c_s^3 \approx 1.5 \times 10^{33} \, n_0 H_{\rm pc} ^2$ erg s$^{-1}$, for an ambient gas at $10^4$ K  and $n_0=1$ cm$^{-3}$. 
For $10^6$ K gas, the requirement is $10^3$ times larger. If we use
$n_0= 10$ cm$^{-3}$, $H_{\rm pc}= 50$, and $T= 10^6$ K, the threshold mechanical luminosity for breakout is $L_{cr}\approx 4 \times 10^{40}$ erg s$^{-1}$.

For a Kroupa/Chabrier IMF, this translates to a threshold SFR of $0.1$ M$_{\odot}$ yr$^{-1}$ in the central region. This also implies a SNe rate of $10^{-3}$ yr$^{-1}$, which can be compared with the SNe rate $2.5 \times 10^{-3}$ yr$^{-1}$ in the nuclear ($35$ pc) region of M82 (Table 1 of \cite{forster03}).  If we considered the whole region of radius $200$ pc and total height $100$ pc, this would imply a low supernova rate density and consequently a small porosity, but that would neglect the effect of coherent SNe inside the small size of a OB association (few tens of pc).  Superbubbles breaking out of a stratified disc ultimately assume an oval shape, with an extent in the plane of the disc  $\approx \pi \times$ the scale height, in the Kompaneets approximation. 
Therefore such a superbubble would ultimately engulf the central region, with radial length scale $\sim 200$ pc and scale height $\sim 50$ pc and the gas inside this region would be shock heated. 

Assuming continuous star formation for the time scale of the wind ($\sim 100$ Myr) and a typical SNe explosion energy of $10^{51}$ erg, this critical luminosity implies a total number  of OB stars of $N_{\rm OB}  
\approx 1.2 \times 10^5 \,
(H_{\rm pc}/50)^2 \, (n_0/10 \, {\rm cm}^{-3})$. For a Kroupa/Chabrier IMF, the corresponding stellar mass is $\approx 1.2 \times 10^7$ M$_{\odot} \, (H_{\rm pc}/50)^2 \, (n_0/10 \, {\rm cm}^{-3})$. \cite{matzner00} have estimated that a fraction $0.25\hbox{--}0.75$ of the molecular cloud mass is ultimately converted into stars. Using the upper limit of $0.75$, one then
finds a conservative estimate for the required total mass of the parent molecular cloud(s) to be $\sim 10^7$ M$_{\odot} \, (H_{\rm pc}/50)^2 \, (n_0/10 \, {\rm cm}^{-3})$. 

Another way of estimating this is to use the empirical SFR in molecular clouds as determined by \citep{lada10}.
They found the SFR to be $4.6\pm 2.6 \times 10^{-8} \, M_{th}$ M$_{\odot}$ yr$^{-1}$, where $M_{th}$ is the mass of the cloud above a density threshold of $n_{th}\approx 10^4$ cm$^{-3}$. The above SFR then implies $M_{th}\approx 2 \times 10^6$ M$_{\odot}$. According to Fig 3 of Lada \etal (2010), such dense clumps amount to 
a fraction $0.03\hbox{--}0.15$ of the total mass of molecular clouds. Using  the upper value of $0.15$ yields a conservative estimate for the  total molecular mass  $M_{th} \sim10^7$ M$_{\odot}$.

For molecular clouds in the central regions of galaxies, the requirement may be stronger than this. In the central regions of galaxies, as in our Galaxy, the turbulent speed is likely to be large. \cite{krumholz05, padoan11}  have shown that due to turbulence, the threshold gas density for star formation increases to $n_{cr}\sim A_x \alpha_{vir} \mathcal{M}^2 \, n$, where $A_x$ is a constant close to unity, $\alpha_{vir}\approx 1.5$ is the virial parameter for the cloud, $n$ is the ambient density, and $\mathcal{M}$ is the Mach number. Typically  $\mathcal{M}\sim 50$, as in the case of the `Brick' cloud in the Galactic CMZ 
\citep{kruijssen13}. For our fiducial MC, with average density $n\approx 5 \times 10^3 \, {\rm cm}^{-3} \, M_{\rm MC, 5} \, R_{\rm MC, 5}^{-3}$, the critical density for star formation then becomes 
$n_{\rm cr}\sim 1.4 \times 10^7$ cm$^{-3}$. We can estimate the mass in clumps with density greater than this by using the fact the probability distribution function of density in a turbulent ISM at high density end is described as a power law, $dp/dn \propto n^{-\gamma}$, with $\gamma\approx 2.5\hbox{--}2.75$ \citep{kritsuk11}. If the cumulative mass (above a certain density) for $n_{\rm th}$ (the threshold without turbulence, $10^4$ cm$^{-3}$) and $n_{\rm cr}$ (the threshold with turbulence) are denoted as $M_{\rm th}$ and $M_{\rm cr}$, then one has $M_{\rm cr}/M_{\rm th}\approx (n_{\rm cr}/n_{\rm th})^{2-\gamma}$. For $\gamma\sim 2.5$, and for the above values of two threshold densities, we then have $M_{\rm cr}\approx (1/40) \, M_{\rm th}$.

Given the uncertainties, we can conclude that the total mass
requirement is increased by a factor $\ge 10$, to $\sim 10^8$ M$_{\odot}$
in order to mitigate the effect of turbulence and create energetic superbubble(s). We can compare this with the observations of molecular mass in the central regions of nearby starburst galaxies. \cite{plenisas97} estimated a molecular mass of $10^8\hbox{--}10^9$ M$_{\odot}$ in the central regions of nuclear starburst galaxies NGC 2903, NGC 3351, NGC 3504. 

This threshold molecular mass  
inside a region of radius $200$ pc in radius implies a
surface density  $\ge 10^{3}$ M$_{\odot}$ pc$^{-2}$, which can be considered  
as a precondition for producing coherent SNe in order to initiate a galactic wind. Since
the Kennicutt-Schmidt SFR is somewhat lower at sub-kpc scale (e.g. \cite{momose13}), 
this surface density implies a SFR of $\sim 1$ M$_{\odot}$ yr$^{-1}$ kpc$^{-2}$ in the nuclear 200 pc region.

For comparison, most of the nearby galaxies observed in the BIMA SONG survey have central CO surface densities less than this and also do not show signs of wind \citep{helfer03}. In contrast, CO observations of ULIRGs show the existence in the central few hundred parsecs region a molecular gas of mass $(0.4\hbox{--}1.5)\times 10^{10}$ M$_{\odot}$, with a surface density of $\sim(0.5\hbox{--}1)\times 10^4$ M$_{\odot}$ pc$^{-2}$ \citep{
solomon97}. We compare the observed molecular surface density of a few nearby starburst with winds in Table 1.

\begin{table}
\caption{Observed central molecular surface density in starbursts with wind [Refs: (1) \cite{aalto12}, (2) \cite{sakamoto11}, (3) \cite{sargent91}, (4) \cite{scoville97}, (5) \cite{sofue01}]
}
\centering
\begin{tabular}{c c c c}\\[0.5ex]
\hline\hline
Name & Size of central region (pc) & $\Sigma_{\rm H_2}$ ($10^3$ M$_{\odot}$ pc$^{-2}$)   \\
\hline
 NGC 1377 & -   & 5$^{(1)}$ \\
 NGC 253 & 300 & 10$^{(2)}$ \\
 Arp 299 &  - & 62 $^{(3)}$ \\
 Arp 220 & 100 & 58 $^{(4)}$ \\
 NGC 3079 & 125 & 100$ ^{(5)}$ \\[1ex]
\hline
\end{tabular}
\label{table:table1}
\end{table}


\section{Discussion}
The nature of the central region of galaxies can affect the mass loading in outflows, especially the
amount of cold material in the outflow. In the case of SNe remnants 
producing a high porosity of hot gas, molecular clouds are likely to be destroyed, whereas in the present scenario, 
the amount of cold/molecular gas in the outflow could be large. This is because 
the time scale of superbubble(s)
 destroying the parent molecular cloud(s) is a few $\times 10$ Myr (lifetime of massive stars), is also 
 the time scale of galactic outflows, and the dynamical effects of starbursts on the surrounding medium. 
 It is therefore reasonable to expect that remnants of molecular clouds will be advected into the outflows in this case,
as has been observed in NGC 253 \citep{bolatto13} with {\it ALMA}, where molecular gas
is seen to envelope the X-ray emitting region and superbubbles can identified. 

It is not necessary that all SNe explode within one OB association in this scenario. The initial trigger can be provided by a super star cluster ($M\sim 10^7$ M$_{\odot}$; e.g. Walcher \etal 2005), after
which the resulting superbubble can enhance star formation in the vicinity. On one hand the shock wave from superbubble can trigger star formation in a nearby cloud, and on the other hand, the increased gas pressure can enhance the SFR \citep{blitz06}. Also, the loss of mass through the superbubble creates pockets of low density gas, which can be heated with high efficiency by latter generation SN remnants and create a reservoir of hot gas.

To conclude, we have argued that it is difficult to thermalize the energy input from SNe in the central regions ($\sim 200$ pc)  of starbursts and create a large filling factor for hot ($\simgt 10^6$ K) gas as is commonly held. We suggested that coherency of SNe is need to create superbubbles that are energetic enough to initiate a galactic wind, and determined a threshold molecular surface density $\simgt 10^3$ M$_{\odot}$ pc$^{-2}$ in the central region for this scenario.

We thank the anonymous referee for useful comments. This work is partly supported by an Indo-Russian project 
(RFBR grant 12-02-92704-IND, DST-India grant INT-RFBR-P121).

\end{document}